\documentclass[a4paper]{jpconf}
\usepackage{graphicx}
\pdfoutput=1
\begin{document}
\title{Prospects for an experiment to measure
  ${\rm BR}(K^0_L\to\pi^0\nu\bar{\nu})$ at the CERN SPS}

\author{M Moulson, for the NA62-KLEVER Project}

\address{INFN Laboratori Nazionali di Frascati, 00044 Frascati RM, Italy}

\ead{moulson@lnf.infn.it}

\begin{abstract}
  Precise measurements of the branching ratios for the
  $K\to\pi\nu\bar{\nu}$ decays can provide unique constraints
  on CKM unitarity and, potentially, evidence for new physics.
  It is important to measure both decay modes, $K^+\to\pi^+\nu\bar{\nu}$
  and $K^0_L\to\pi^0\nu\bar{\nu}$, since different new physics models affect
  the rates for each channel differently. We are investigating the
  feasibility of performing a measurement of BR($K^0_L\to\pi^0\nu\bar{\nu}$)
  using a high-energy secondary neutral beam at the CERN SPS in a successor
  experiment to NA62. The planned experiment would reuse some of the NA62
  infrastructure, including possibly the NA48 liquid-krypton calorimeter.
  The mean momentum of $K^0_L$ mesons decaying in the fiducial volume is 70 GeV;
  the decay products are boosted forward, so that less demanding
  performance is required from the large-angle photon veto detectors.
  On the other hand, the layout poses particular challenges for the
  design of the small-angle vetoes, which must reject photons from $K^0_L$
  decays escaping through the beam pipe amidst an intense background from
  soft photons and neutrons in the beam. We present some preliminary
  conclusions from our feasibility studies, summarizing the
  design challenges faced and the sensitivity obtainable for the
  measurement of BR($K^0_L\to\pi^0\nu\bar{\nu}$).
\end{abstract}

\section{Introduction}

\begin{figure}[ht]
\centering
\includegraphics[width=100mm]{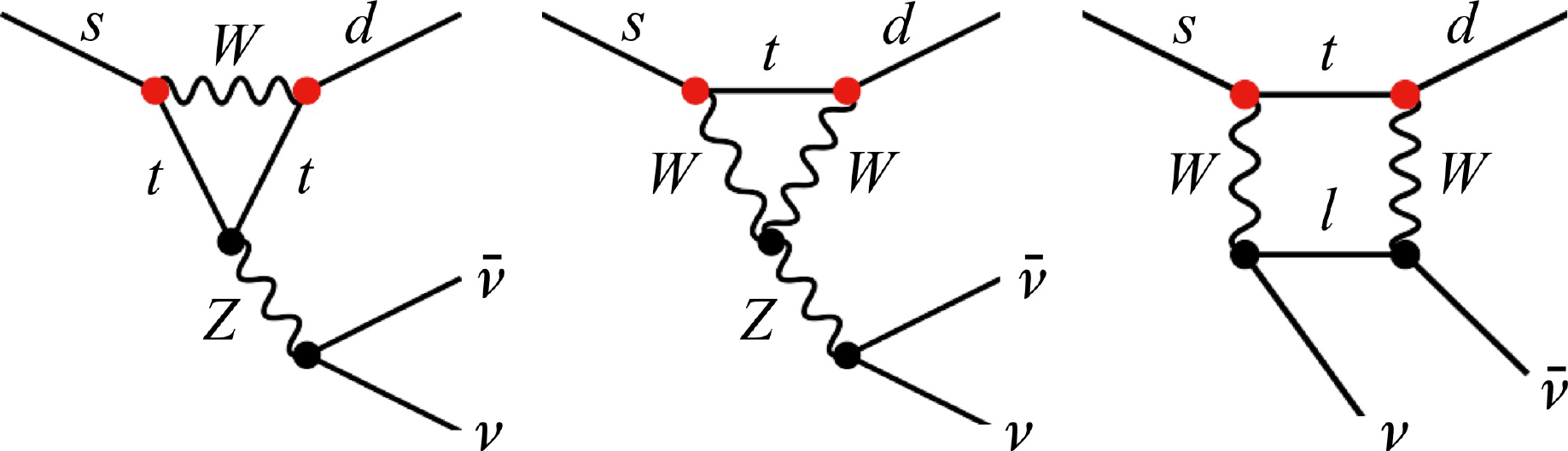}
\caption{Diagrams contributing to the process $K\to\pi\nu\bar{\nu}$.} 
\label{fig:fcnc}
\end{figure}

The $K\to\pi\nu\bar{\nu}$ decays are flavor-changing neutral current (FCNC)
processes that probe the $s\to d\nu\bar{\nu}$ transition via the 
$Z$-penguin and box diagrams shown in figure~\ref{fig:fcnc}. They are 
highly GIM suppressed and their Standard Model (SM) rates are very small.
For several reasons, the SM calculation for their branching ratios
(BRs) is particularly clean (see \cite{C+12:kaonRev} for a review):
\begin{itemize}
\item The loop amplitudes are dominated by the top-quark contributions.
The neutral decay violates $CP$; its amplitude involves the top-quark 
contribution only.
\item The hadronic matrix element for these decays can be obtained from the
precise experimental measurement of the $K_{e3}$ rate.
\item There are no long-distance contributions from processes with
intermediate photons. 
\end{itemize}
In the SM,
${\rm BR}(K^+\to\pi^+\nu\bar{\nu}) = (8.4 \pm 1.0)\times10^{-11}$ and
${\rm BR}(K^0_L\to\pi^0\nu\bar{\nu}) = (3.4 \pm 0.6)\times10^{-11}$
\cite{B+15:KpnnSM}.
The uncertainties are entirely dominated by the CKM inputs, which in
this case are from tree-level observables. Without the parametric errors
from this source, the uncertainties would be just $0.30\times10^{-11}$ (3.5\%)
and $0.05\times10^{-11}$ (1.5\%), respectively. Because of corrections
for lighter-quark contributions to the amplitudes, the intrinsic
uncertainty is slightly larger for the charged channel. 

\begin{figure}[ht]
\centering
\includegraphics[width=80mm]{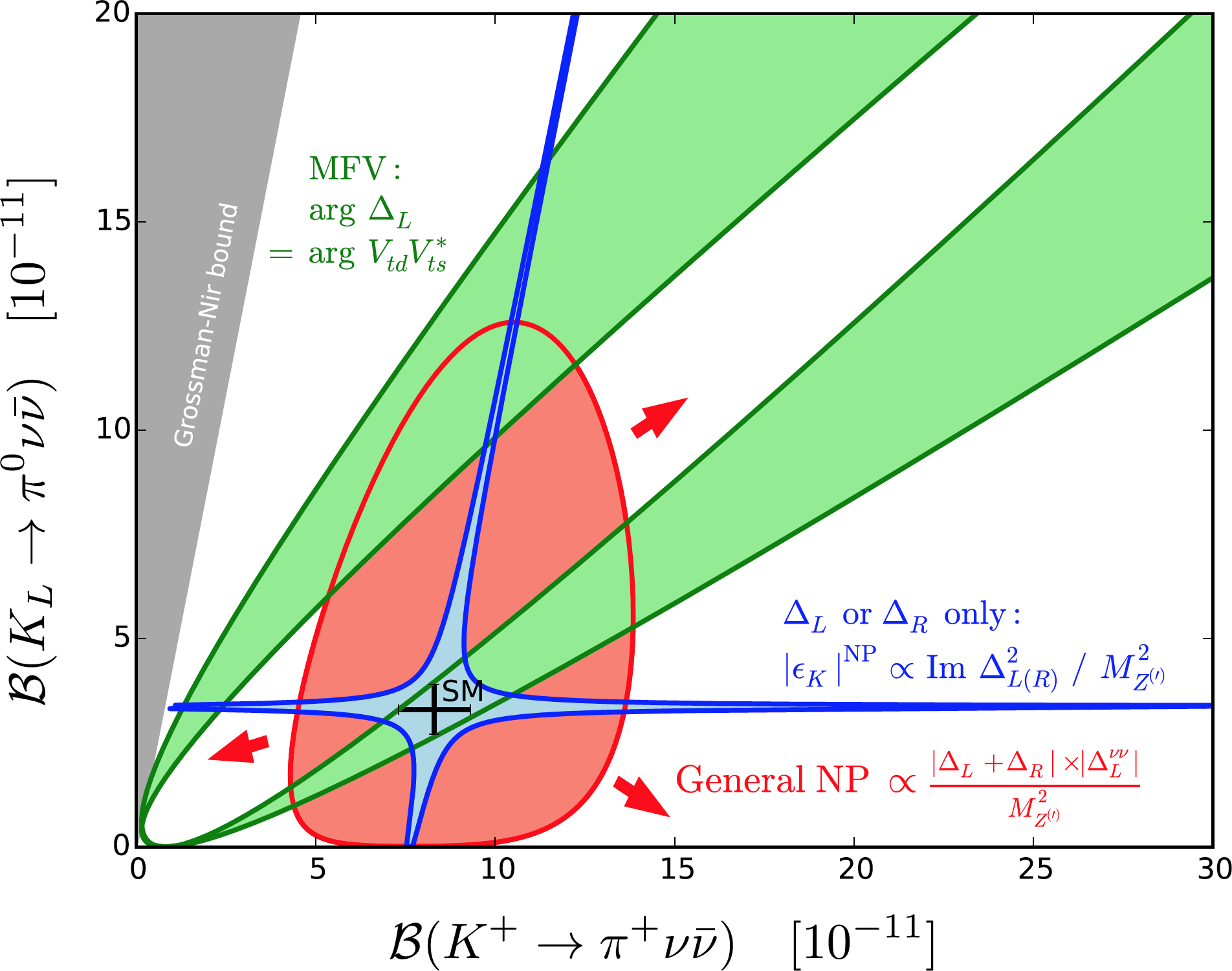}
\caption{Schematic illustration of correlations between BRs for
  $K^+\to\pi^+\nu\bar{\nu}$ and $K^0_L\to\pi^0\nu\bar{\nu}$ expected
  in different new-physics scenarios, from \cite{BBK15:KpnnBSM}.}
\label{fig:bsm}
\end{figure}
Because the SM rates are small and predicted very precisely,
the BRs for these decays are sensitive probes for new physics. 
In general, ${\rm BR}(K^0_L\to\pi^0\nu\bar{\nu})$ and
${\rm BR}(K^+\to\pi^+\nu\bar{\nu})$ are differently sensitive to
modifications from a given new-physics scenario.
If one or both BRs is found to differ from its SM value, it may be possible
to characterize the physical mechanism responsible, as schematized in
Figure~\ref{fig:bsm}, from \cite{BBK15:KpnnBSM}.
For example, if the pattern of flavor-symmetry breaking from new physics
were the same as in the SM (minimal flavor violation), the $K^0_L$ and $K^+$
BRs would lie along the band of correlation shown in green.
If the new interaction were to couple to only left-handed or only
right-handed quark currents, as expected for example in models with
modified $Z$ couplings or littlest Higgs models with $T$ parity, the
BRs would lie along one of the branches shown in blue.
New physics without these constraints, as expected for example
in models with large extra dimensions, could modify the $K^+$ and $K^0_L$ BRs
in an arbitrary way, as illustrated in red.

The BR for the decay $K^+\to\pi^+\nu\bar{\nu}$ has been measured by 
Brookhaven experiment E787 and its successor, E949. The combined result
from the two generations of the experiment, obtained with seven candidate 
events, is ${\rm BR}(K^+\to\pi^+\nu\bar{\nu}) = 
1.73^{+1.15}_{-1.05}\times10^{-10}$ \cite{E949+09:Kpnn2}. 
The goal of the NA62 experiment at the CERN SPS~\cite{NA62+10:TDD} is
to measure ${\rm BR}(K^+\to\pi^+\nu\bar{\nu})$ with a precision of about 10\%.
NA62 is currently running and is expected to collect $\sim100$ signal
events by the end 2018. 

The decay ${\rm BR}(K^0_L\to\pi^0\nu\bar{\nu})$ has never been measured.
The KOTO experiment at J-PARC has a good chance 
of observing it. The experiment makes use of a tightly collimated,
low-energy neutral beam (peak momentum 1.4~GeV) and compact, hermetic
detector. From a brief pilot run in 2013, KOTO obtained the limit
${\rm BR}(K^0_L\to\pi^0\nu\bar{\nu}) < 5.1\times10^{-8}$
(90\% CL)~\cite{KOTO+16:Kpnn13}. The experiment has been running since
2015. From the preliminary analysis of 5\% of the data collected so far,
KOTO has reached a single-event BR sensitivity of $5.9\times10^{-9}$;
background levels are still under evaluation~\cite{KOTO+16:Kaon}.
By the end of 2015, the beam power reached 42~kW; it is expected to gradually
increase to 100~kW. If this can be done, the experiment should reach
single-event sensitivity for the SM BR by about 2019. In the longer term,
KOTO strongly intends to upgrade the beam and experiment to perform a
measurement with $\sim$100 event sensitivity.  A plan to do
so was outlined in the original 2006 KOTO proposal, but there is no
official Step 2 proposal yet. The Step 2 measurement would require
construction of a new neutral beamline, a complete rebuild of the detector,
and extension of the experimental hall; data taking would not be expected
to begin before 2025.

\section{KLEVER}

Given the both the importance and the difficulty of the measurement
of ${\rm BR}(K^0_L\to\pi^0\nu\bar{\nu})$, an experiment
making use of a technique complementary to that used by KOTO
is well motivated.
We are evaluating the feasibility of an experiment at the CERN SPS to
measure ${\rm BR}(K^0_L\to\pi^0\nu\bar{\nu})$ using a high-energy beam.
This makes photon vetoing significantly easier, but increases considerably
the size of the detector, and in particular, the volume to be covered with
photon vetoes. On the other hand, since the photons from background $K^0_L$
are boosted forwards, the coverage of the large-angle photon vetoes may
not need to extend beyond 100~mrad in the polar angle.
The experiment could reuse some of the NA62 experimental infrastructure,
and possibly the NA48 liquid-krypton calorimeter~\cite{NA48+07:NIM}. 
Given the time needed for R\&D and construction, in addition to the
constraints from the LHC running schedule, the natural target date for
the experiment to turn on would be at the start of LHC Run 4, in early
2026.

\subsection{Beam}

Assuming the SM BR for $K^0_L\to\pi^0\nu\bar{\nu}$ and an acceptance
of 10\% for decays within the experiment's fiducial volume (FV),
$3\times10^{13}$ $K^0_L$ decays would be required for the observation of
100 signal events. Extraction of the 400~GeV proton beam onto a
$1\lambda_{\rm int}$ beryllium target to produce a tightly collimated
(0.28~$\mu$sr) neutral beam at an
angle of 2.4~mrad (chosen to optimize the ratio of $K^0_L$ decays in the
FV to neutrons in the beam) would result in a $K^0_L$ flux of
$2.8\times10^{-5}$ per proton on target (pot).
About 2.2\% of the $K^0_L$ mesons in the beam would decay in a 50-m FV
with an average momentum of 70 GeV. The required integrated proton flux
is then $5\times10^{19}$ pot. Assuming that this is delivered in
5 years\footnote{100 effective days of running per year, similar to a
  Snowmass year.}
the required primary beam intensity would be $2\times10^{13}$ protons per
pulse (ppp) with one pulse every 16.8~s. The sixfold increase in
intensity relative to NA62 is made necessary by the tight beam collimation
and long $K^0_L$ lifetime.

The maximum intensity that the SPS can deliver to the
North Area is about $4\times10^{13}$ ppp, but apart from competition for
protons from the SPS, a primary beam intensity of $2\times10^{13}$ ppp is not
currently available on any of the North Area targets.
Running with an intensity this high would require comprehensive upgrades
to the current beamline cavern and experimental area, as well as the nearly
2 km long beam transport from the extraction point in the SPS to the
production target of the experiment. An alternative could be to site the
experiment at the North Area Beam Dump Facility under discussion in the
context of the SHiP experiment. Either way, detailed solutions and meaningful
cost estimates will require studies in collaboration with the CERN Accelerator
and Technology Sector.

\begin{figure}
\begin{center}
\includegraphics[width=90mm]{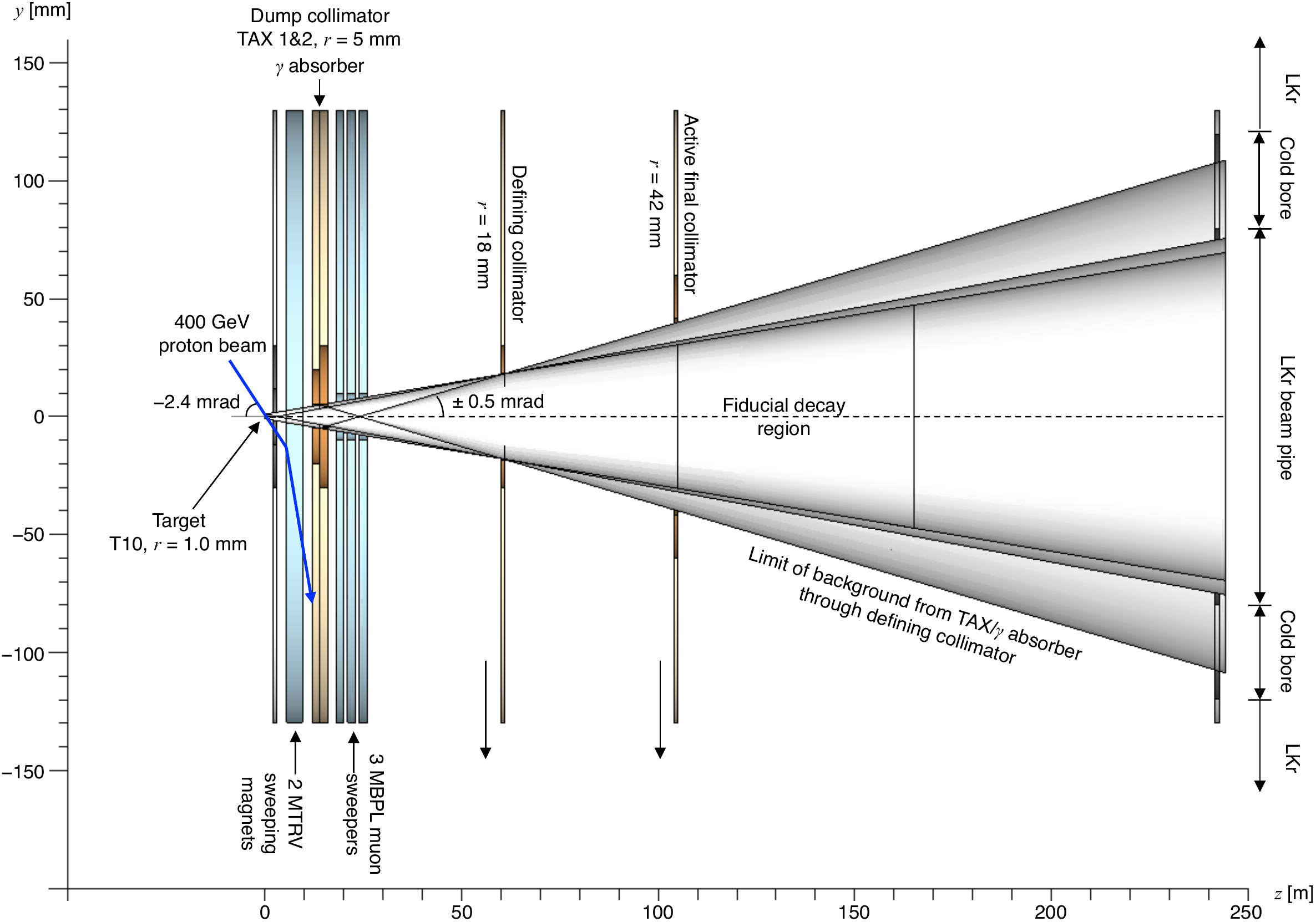}
\end{center}
\caption{Layout of the neutral beamline.}
\label{fig:beam}
\end{figure}
The layout of the secondary beamline is illustrated in figure~\ref{fig:beam}.
The 400-GeV beam is incident on the target at a downward angle of 2.4~mrad.
Three collimators are used to obtain a neutral beam with an opening angle
of 0.3~mrad: a moveable dump collimator at $z=15$~m from the target,
the defining collimator at $z=60$~m, and a final collimator at $z=105$~m
with an aperture just larger than the angular limit of the background from
the dump collimator, to remove particles showering on the edges of the
defining collimator.
The 45~m length between the defining and final collimators is equivalent to
nearly 9 $K^0_S$ lifetimes at the mean beam momentum.
Vertical sweeping magnets are located just upstream of the dump collimator,
and horizontal muon-sweeping magnets are just downstream. A photon converter
of high-$Z$ material is placed in the center of the dump collimator, to
reduce the flux of photons in the neutral beam.

The $K^0_L$, photon, and neutron fluxes in this beamline have been simulated
with Fluka and Geant4, with the former used for the beam-target interaction
and the latter used for propagation of secondaries through the beamline.
The simulation predicts 290~MHz of $K^0_L$ in the beam, about 50\% more than
the parameterizations used to estimate the intensity requirement would suggest.
The simulation also predicts 230~MHz of photons with $E>5~{\rm GeV}$
and 20~MHz of photons with $E>30{\rm~GeV}$, assuming a 30-mm iridium absorber
in the dump collimator. More problematic is the fact that the simulation
predicts 3~GHz of neutrons in the beam, which places stringent requirements
on the insensitivity of the small-angle calorimeter to neutrons. 

\subsection{Detector}

The experimental signature of the $K^0_L\to\pi^0\nu\bar{\nu}$ decay
consists of two photons with unbalanced transverse momentum in an event
in which there are no additional particles. The only available kinematic
constraint is on the two-photon invariant mass: $M(\gamma\gamma) = m_\pi^0$.
The most dangerous background is from $K^0_L\to\pi^0\pi^0$ decays (which
are about $3\times10^7$ times more abundant than the signal) with two
lost photons.
The centerpiece of the experiment is the NA48 liquid-krypton calorimeter,
which is used both to reconstruct the $\pi^0$ for the signal decay and to
veto any other photons present. The invariant mass constraint
is imposed to obtain the $z$ position of the $\gamma\gamma$ vertex.
\begin{figure}
\begin{center}
\includegraphics[width=120mm]{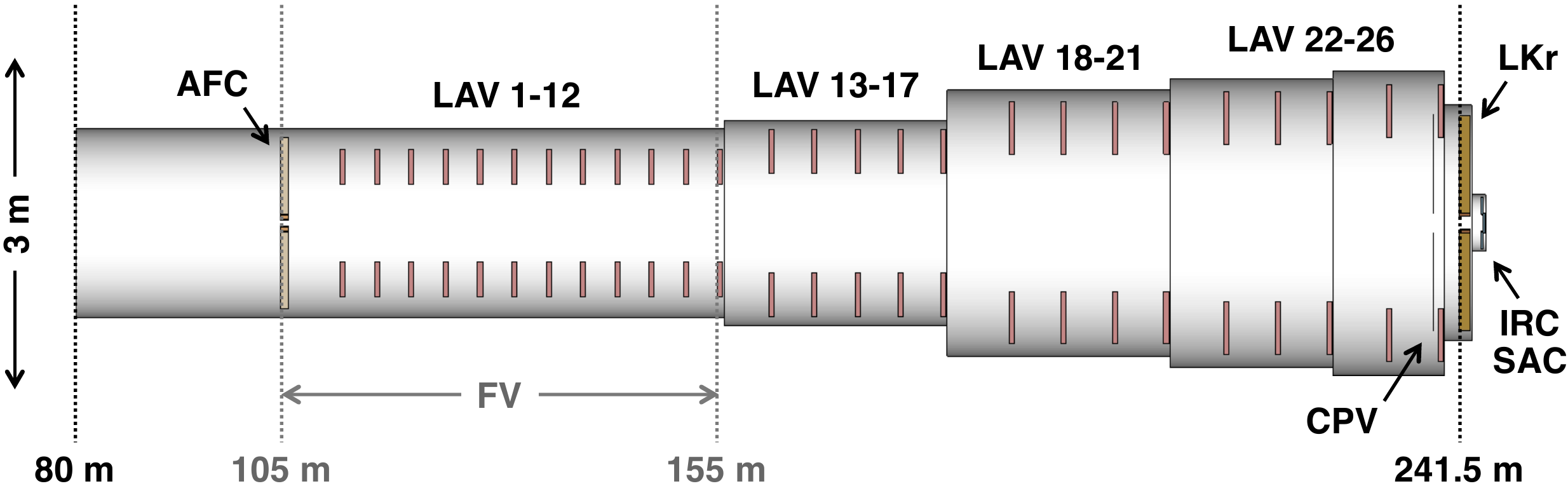}
\end{center}
\caption{Schematic illustration of the experimental layout. The scales for
  transverse and longitudinal dimensions are in the ratio 10:1.}
\label{fig:layout}
\end{figure}
The layout of the detector elements is schematically illustrated in
figure~\ref{fig:layout}. The largest elements are about 3 m in diameter.

The vacuum volume begins 25~m upstream of the final collimator.
Decays in this region for which two photons pass through the final
collimator can give rise to background.
The final collimator itself is thus an active detector, with the
collimating surfaces made of LYSO surrounded by an electromagnetic
calorimeter extending out to a radius of 1~m to provide
upstream veto coverage.

The FV extends from $z = 105~{\rm m}$ (the position of the
final collimator) to $z = 155~{\rm m}$. The front face of
the sensitive part of the LKr calorimeter is at z = 241.5 m.
The 90-m distance from the downstream end of the FV to the LKr
significantly aids with background rejection, since most $K^0_L\to\pi^0\pi^0$
decays with lost photons occur just upstream of the LKr. Even
for $K^0_L\to\pi^0\pi^0$ events in which the two detected photons are
from different $\pi^0$s, the vertex is almost always reconstructed
downstream of the FV.

Besides the LKr, there are a total of 26 ring-shaped large-angle photon veto
(LAV) stations in five different sizes, placed at intervals of 4 to 6 m to
guarantee coverage out to $\theta = 100~{\rm mrad}$. The LAV detectors
are based on the lead/scintillating-tile design of the large-angle photon
vetoes for the (canceled) CKM experiment at Fermilab~\cite{RCT04:VVS}.

The small-angle vetoes (IRC and SAC) on
the downstream side of the LKr intercept photons from $K^0_L$ decays that pass
through the beam pipe.
The small-angle calorimeter (SAC) squarely intercepts
the neutral beam.
The intermediate-ring calorimeter (IRC) is a ring-shaped
detector between the SAC and LKr that intercepts photons from downstream
decays that make it through the calorimeter bore at slightly larger angles.
Because of the high rates of neutrons and photons in the
beam, the design of these detectors, in particular the SAC, is one of the
most challenging aspects of
the experiment. The task is made a little easier by the fact that photons
from $K^0_L$ decays in the SAC acceptance have very high energy. The SAC can be
blind to photons with $E < 5~{\rm GeV}$, and need only have very high
efficiency (99.99\%) for photons with $E > 30~{\rm GeV}$. The current
design calls for a compact tungsten/silicon-pad sampling calorimeter
with a crystal metal absorber. The coherent interaction of
photons with the crystal lattice reduces the radiation length of the absorber,
and hence the ratio $X_0/\lambda_{\rm int}$ (see, e.g.,~\cite{B+88:pair}),
resulting in a calorimeter with better transparency to neutrons.
The effect is greatest for high-energy photons; for 30~GeV photons incident
at an angle of 2~mrad, the pair-production cross section is enhanced by a
factor of about three.
Preliminary Geant4 simulations indicate that perhaps 20\% of the neutrons
in the beam will interact in the SAC; information on the transverse and
longitudinal shower spread and pulse shape information will be used to
obtain additional neutron discrimination power.

In addition to the photon vetoes, the experiment makes use of
charged-particle veto (CPV) detectors and a hadronic calorimeter downstream
of the LKr (not shown in figure~\ref{fig:layout}), to reject background
from the copious $K^0_L$ decays into charged particles.

\subsection{Expected performance}

\begin{figure}
\begin{center}
\includegraphics[width=120mm]{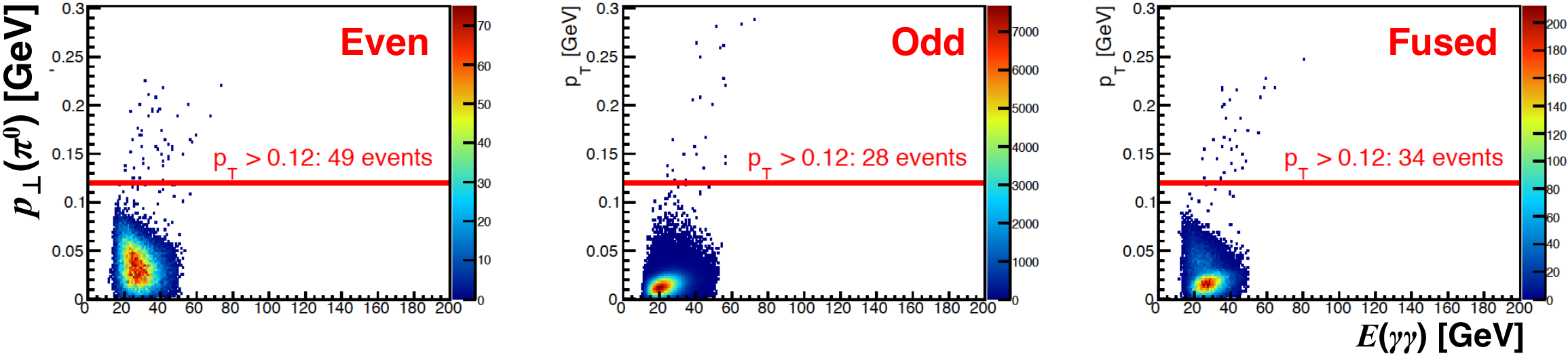}
\end{center}
\caption{Distribution in $(p_{\perp\,\gamma\gamma},E_{\gamma\gamma})$ plane
  of $K^0_L\to\pi^0\pi^0$ events passing all selection cuts, from fast
  simulation: even events (left), odd events (center), and events with fused
  clusters (right).}
\label{fig:bkg}
\end{figure}
The rejection power for background from $K^0_L\to\pi^0\pi^0$ events has been
studied using a fast Monte Carlo, with statistics equivalent to those from
five years of data taking. Signal candidates
are required to have exactly two clusters on the LKr and no hits on any other
detector. The decay vertex position in $z$ is obtained from the separation
between the clusters on the calorimeter and the cluster energies, and is
required to be inside the FV.
There is a substantial amount of background from upstream decays, but this
can be eliminated by requiring that the clusters on the LKr lie outside of
a radius of 35~cm from the beam axis. Figure~\ref{fig:bkg} shows the
distribution of the remaining events in the plane of pair transverse momentum
$p_{\perp\,\gamma\gamma}$ vs.\ pair energy $E_{\gamma\gamma}$ for events in
which  both photons come from the same $\pi^0$ (even pairing), different
$\pi^0$s (odd pairing), or in which one or both clusters contain
overlapping LKr hits (fused). If $p_{\perp\,\gamma\gamma}$ for signal
candidates is required to be larger than 0.12~GeV, 111 background events
remain. Similar high-statistics studies of
$K^0_L\to\pi^0\pi^0\pi^0$ and $K^0_L\to\gamma\gamma$ backgrounds
give 15 and 0 counts, respectively ($K^0_L\to\gamma\gamma$ is effectively
eliminated by the $p_\perp$ cut).
The acceptance for $K^0_L\to\pi^0\nu\bar{\nu}$ decays in the fiducial volume is
about 9\% (the $p_{\perp\,\gamma\gamma}$ cut retains 78\% of signal events).
In five years, 97 signal events are expected to be collected.
The photon converter in the dump collimator may absorb or scatter
as much as 35\% of the beam. This effect is not simulated; taking it into
account, at the SM BR, about 60 signal events per year are expected, together
with a similar number of background events ($S/B \sim 1$).

\section{Conclusions}

The present results are highly preliminary. In particular, high-statistics
background studies have been carried out only for a few channels.
Some potential backgrounds, such $nn\to nn\pi^0$ from interactions of beam
neutrons on residual gas, need to be investigated. The detector concepts
require validation, and the suitability of the LKr performance, in particular
concerning photon detection efficiency and time resolution, needs to be
verified.
The possibility of adding charged-particle tracking to the experiment,
which would provide more complete final-state reconstruction for efficiency
estimation and systematic control, and which would allow significant
expansion of the physics program, is under investigation.
Finally, the available intensity and siting options at CERN need to be worked
out. With these caveats, our preliminary design studies indicate that an
experiment to measure ${\rm BR}(K^0_L\to\pi^0\nu\bar{\nu})$ can be performed
at the SPS during LHC Run 4 (2026-2029). These ideas are under development
in the context of the CERN Physics Beyond Colliders initiative~\cite{Web:PBC},
with the intention to move towards an official proposal within the next
two years. 

\section*{References}
\bibliographystyle{iopart-num}
\bibliography{kaon16_proc}

\end{document}